# How does light move?
## Determining the flow of light without destroying interference

M. Davidović[1] and A. S. Sanz[2]

[1] Faculty of Civil Engineering, University of Belgrade – Belgrade, Serbia

[2] Instituto de Física Fundamental, Consejo Superior de Investigaciones Científicas – Madrid, Spain

**Young's two-slit experiment constitutes the paradigm of quantum complementarity. According to the complementarity principle, complementary aspects of quantum systems cannot be measured at the same time by the same experiment. This has been a long debate in quantum mechanics since its inception. But, is this a true constraint? In 2011, an astounding realization of this experiment showed that perhaps this is not the case and the boundaries to our understanding of the quantum world are still far away.**

210 years ago Thomas Young presented the outcomes of his nowadays world-renowned two-slit experiment to the Royal Society [1]. This experiment confirmed Huygens' wave theory of light and refuted Newton's corpuscular one. About 100 years later, though, Albert Einstein again suggested that light was composed by quanta of electromagnetic radiation or photons. Ever since, Young's experiment has constituted the simplest and most elegant proof of the fuzzy dual behavior displayed by quantum systems, both light (photons) and material particles (electrons, atoms, molecules, etc.). Depending on how the experiment is performed, a different complementary aspect of the system is revealed: wave or corpuscle.

In 2011, Aephraim Steinberg and colleagues from the University of Toronto caused a stir in the physics community [2,3] with their challenging realization of Young's experiment with single photons [4]. As they showed, in a certain sense, going beyond the restrictions imposed by both Bohr's complementarity principle and Heisenberg's uncertainty principle is actually feasible. From measurements of the photons' transversal momentum, this group was able to determine the energy streamlines associated with the photon electromagnetic field and, therefore, to infer "which"-slit information (corpuscle behavior) without destroying the interference pattern (wave behavior). In other words, these measurements imply that the field contributing to each half of the interference pattern comes uniquely from the slit that is placed just in front of it, thus indirectly revealing *which* slit the field (photons) passed through. But, how is this possible? Is it not in contradiction with our standard perception and understanding of the quantum world?

### *What are Bohmian trajectories?*
Leaving aside conceptual issues connected to the complementarity and uncertainty principles, there is no reason that impedes us to formulate models aimed at locally tracking the evolution of quantum systems. Actually, from a practical viewpoint, such models are very convenient: they provide us with a feeling of how the probability flows in (configuration) space and time. This is the case of Bohmian mechanics [5], a hydrodynamic formulation of quantum mechanics, where the evolution of quantum systems is described in terms of streamlines or trajectories. This is possible, because this approach focuses on the phase information encoded in the wave function. Thus, for example, in interference phenomena, bundles of trajectories gather along certain directions (maxima), while avoid others (minima).

Bohmian mechanics is applicable whenever the quantum system is described by Schrödinger's equation. But, what happens if we are dealing with light instead? The answer is simple. Given that light interference patterns arise from the accumulation of a large number of photons [6], they can be well described by standard (classical) electromagnetism. Following the Bohmian prescription, an analogous model can then be formulated, where the trajectories (electromagnetic energy flow lines) are obtained from the Poynting vector [7,8] and describe the spatial distribution of the electromagnetic energy density.

### *Weak measurements vs strong or von Neumann measurements*

Now, is it possible to experimentally test the feasibility of the above model? Appealing to the complementarity and uncertainty principles, the immediate answer is "not". Standard (strong) measurements do not follow a unitary evolution transformation, inducing an irreversible change in the system evolution. This problem can be overcome, though, by performing "weak measurements" [9]. These are tiny perturbations performed on the system, which do not alter much its evolution, but that, when averaged on a large number, render complementary information about it. In practice, these measures are equivalent to transition probabilities between two different states, the transition being described by a certain operator. If this operator corresponds to the momentum operator, the average coincides with the Bohmian momentum [10]. In other words, a weak measurement is just a measure of the local flow of the probability density or, equivalently, the local value of the quantum probability current density, often regarded as a non-observable. In the case of light, this translates into a local measure of the photon transversal momentum. This momentum, when averaged over many photons, happens to be proportional to the transversal component of the Poynting vector.

### *Measuring average photon paths experimentally*

In the experiment [4] (see box), single photons produced by a quantum dot pass through a 50:50 beam splitter, which plays the role of Young's two slits. These photons are prepared in a diagonal polarization state, after which they pass through a thin chip of calcite, where the weak measurement is performed: the inclination of the calcite optical axis rotates the photon state, which becomes slightly elliptically polarized. By means of a quarter-wave plate (QWP) and a polarization beam displacer, the two polarization components are eventually separated (according to a circular polarization basis set), each one giving rise to an interference pattern. The shift between these two patterns is proportional to the photon transversal momentum at a particular position --- in other words, from the intensities of the left-hand and right-hand circular components, the weak value of the photon transversal momentum is extracted. Averaging over a large number of photons on that position, not only the typical fringe interference pattern is reconstructed, but also the photon transversal momentum distribution. In order to obtain information at different distances from the "two slits", a set of three lenses is used: by displacing one of them (the middle one), one achieves the effect of detecting the photons at imaging planes closer to or further away from two slits. Experimental results at four different imaging planes are shown in Figs 1a-d.

The information provided by a sequence of transversal momentum distributions recorded for many consecutive and closely spaced imaging planes is then used to determine the average flow of photons. It is here where Bohmian mechanics comes into play. Actually, to be more precise, Bohmian mechanics provides the idea and classical electromagnetism the theoretical framework, as mentioned above. The corresponding trajectories are reconstructed by propagating a set of initial conditions with the aid of the momenta along the transversal direction (see bottom panel of Fig. 1).

In spite of the complexity involved in the experimental setup, the trajectories themselves are a result that can be easily explained in terms of classical electromagnetism. To understand this basic idea, consider two slits such that, when they are illuminated by a monochromatic light, they produce two diffracted Gaussian beams [11]. The energy density of the electromagnetic field behind the slits distributes as shown in Fig. 2a, while its phase is as displayed below, in Fig. 2b. The relation between the corresponding Poynting vector and the electromagnetic energy density gives a velocity field, which accounts for the local transport of energy. The photon transversal momenta (weak values) of Fig. 1 correspond to the transversal components of this field. This correspondence can be seen in Fig. 2c together with some trajectories. This quantity is compared with the experimental data at different distances from the two slits and using different initial electromagnetic energy density distributions [11] (Gaussian and non-Gaussian).

## *Plato and photon paths*

In Plato's Allegory of the Cave, a series of people are enforced to face a wall where they observe the projected shadows of some objects passing by behind them. For those people, these shadows constitute their closer notion to the idea of reality, without ever knowing what the true nature of the objects that cause such shadows is. To some extent, the quantum world operates in a similar fashion: we can only understand quantum systems in a rather limited way. Under these circumstances, the Bohmian formulation offers us the possibility to locally describe the evolution of quantum systems in terms of well-defined trajectories in the configuration space and time. These trajectories, in compliance with the (global) evolution accounted for by the wave function, are not in contradiction, though, with the complementarity and uncertainty principles (understood in a broader sense than it is commonly done). This is an appealing idea from which a richer picture of the physical nature of quantum systems can be extracted, as the above experiment or some other that are currently being proposed [12,13] show.

In that sense, even though the trajectories reconstructed from the experiment cannot be associated with the paths followed by individual photons, but with electromagnetic energy streamlines, the experiment constitutes an important milestone in modern physics. The fact that the trajectories do not cross mean that, at the level of the average electromagnetic field (or the wave function, in the case of material particles, in general), full which-way information can still be inferred without destroying the interference pattern. That is, rather than complementarity, the experiment seem to suggest that superposition has a tangible (measurable) physical reality [14], in agreement with a recent theorem on the realistic nature of the wave function [15].

**BOX: WEAK MEASUREMENTS IN YOUNG'S TWO SLITS**

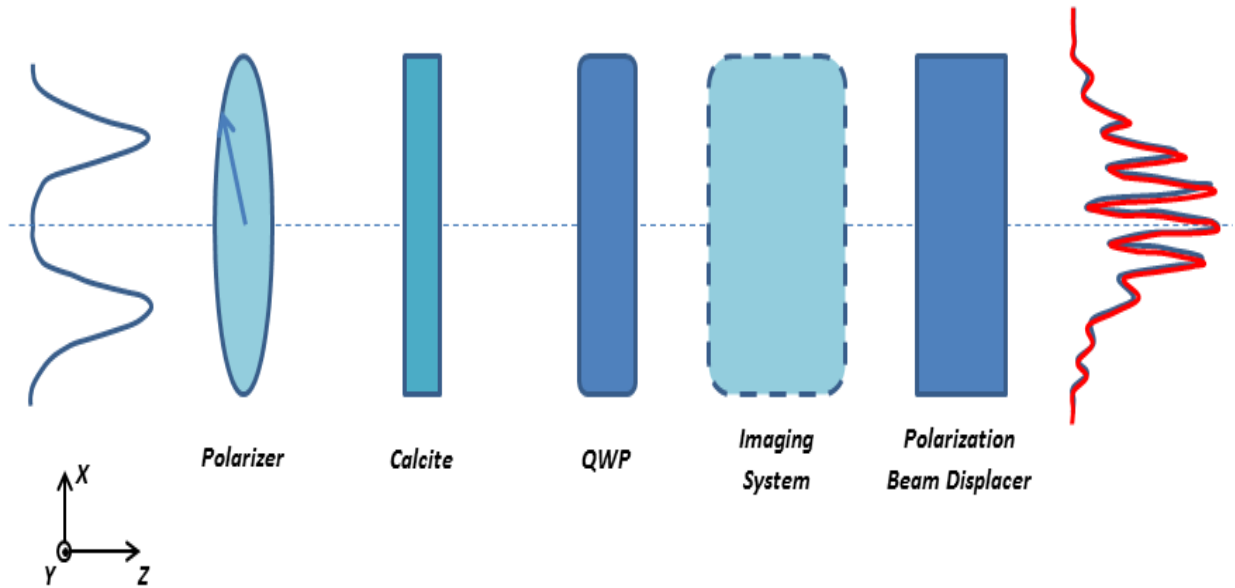

Simplified scheme of the version of Young's experiment prepared by Kocsis et al. [4]. Photons are prepared with diagonal polarization $|D\rangle \propto |H\rangle + |V\rangle$ when they cross the first polarizer. A thin calcite chip with optical axis at 42° induces a small phase-shift between the two photon polarization components (weak measurement), which is a linear function of the photon transverse momentum, $\varphi(k_x) = \zeta(k_x/k) + \varphi_0$ (in the experiment the calcite is tilted in such a way that $\varphi_0 = 0$). The photon polarization state then becomes $|\Psi\rangle \propto e^{-i\varphi(k_x)/2}|H\rangle + e^{i\varphi(k_x)/2}|V\rangle$, which can also be recast as $|\Psi\rangle \propto \left(e^{-i\varphi(k_x)/2} + ie^{i\varphi(k_x)/2}\right)|R\rangle + \left(e^{-i\varphi(k_x)/2} - ie^{i\varphi(k_x)/2}\right)|L\rangle$, in the circularly polarized basis set, $|H\rangle = (|R\rangle + |L\rangle)/\sqrt{2}$ and $|V\rangle = (|R\rangle - |L\rangle)/\sqrt{2}$. These two polarization components give rise to two separate and independently detected interference patterns (strong measurement), with intensities $I_R \propto 1 - \sin\varphi(k_x)$ and $I_L \propto 1 + \sin\varphi(k_x)$. The phase-shift is obtained from the relation $\sin\varphi(k_x) = (I_L - I_R)/(I_L + I_R)$, which relates to the transversal momentum as $k_x = (k/\zeta)\arcsin\{(I_L - I_R)/(I_L + I_R)\}$.

**FIG. 1.** From (a) to (d), experimental intensities (read and blue curves) for the two circular components obtained from photon counts on a CCD camera, and weak momentum values obtained from these intensities at different imaging planes [2]. (e) Reconstruction of average photon (Bohmian) trajectories from weak momentum values taken on 41 imaging planes covering a range of 2.75 to 8.2 meters (the vertical red dashed lines denote the position of the imaging planes of the measures shown in the above panels). Results obtained from Ref. [2]. Reprinted with permission from AAAS.

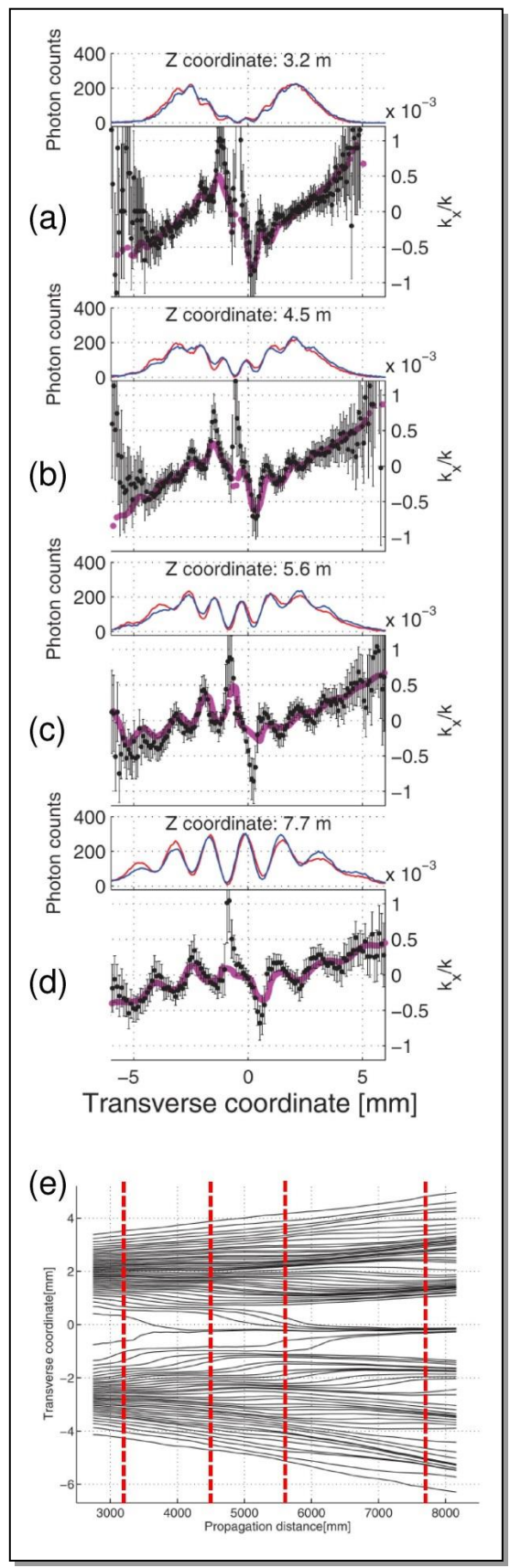

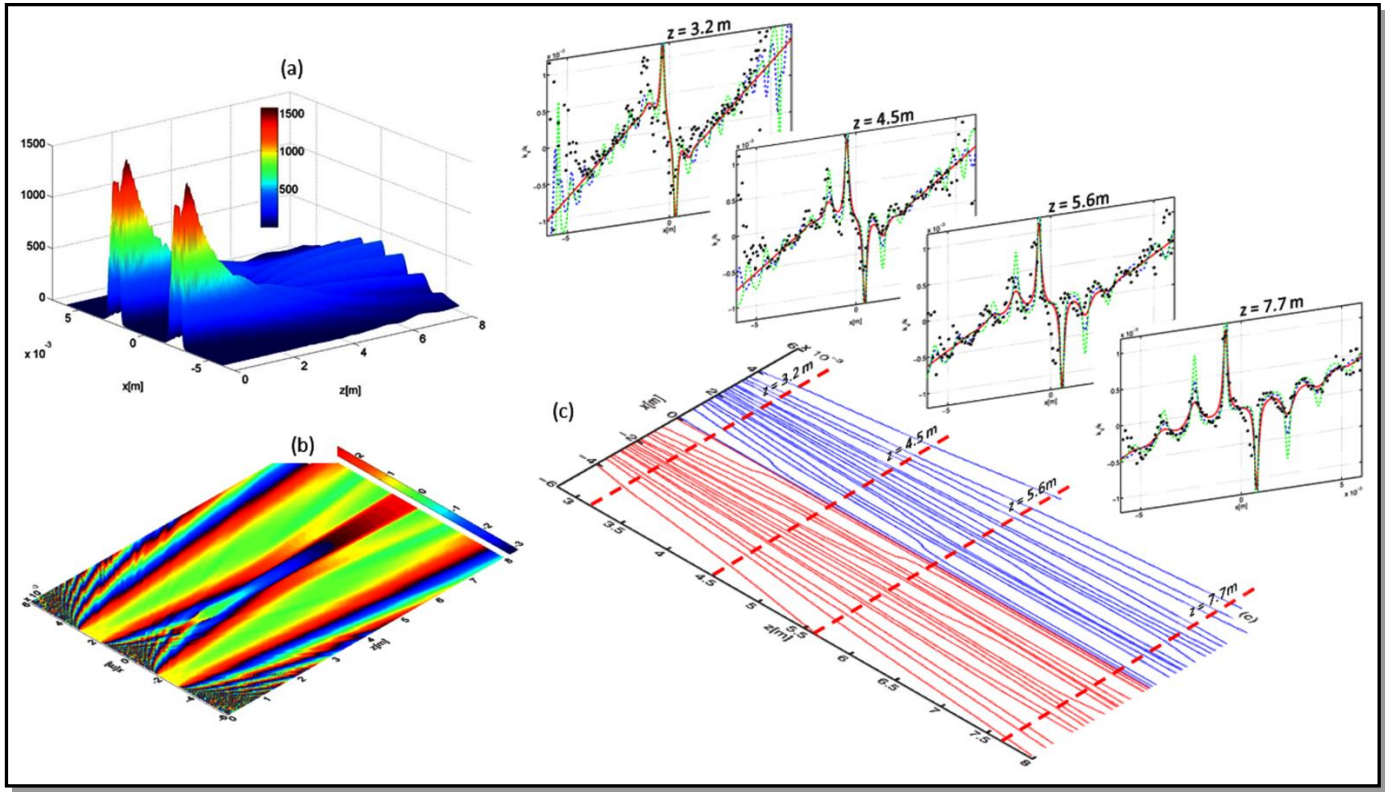

**FIG. 2.** Young's two slit experiment with light. Energy density distribution (a) and phase field (b) of the electromagnetic field generated by two identical, Gaussian slits [11].The corresponding electromagnetic energy flow lines or average (Bohmian) photon trajectories are displayed in (c), together with a comparison between transversal momenta obtained from numerical simulations with different models of slits (colored curves) and the corresponding sets of experimental data [2] (full circles).